\pdfoutput=1
\documentclass[sigconf,nonacm]{acmart}

\acmYear{2020}
\acmISBN{} 
\acmDOI{} 
\startPage{1}

\setcopyright{none}

\usepackage[capitalize]{cleveref}
\usepackage{colortbl}
\usepackage{graphicx}
\usepackage{listings}
\usepackage{natbib-line-breaks}
\usepackage{xfrac}
\usepackage{refs}
\usepackage{rqbox}
\usepackage{soul}
\usepackage{tikz}
\usepackage{tikz-emph}
\usepackage{adjustbox}
\usetikzlibrary{
  calc,
  positioning,
  shapes,
}
\usepackage{pgfplots}
\usepackage{pgfplotstable}
\usepackage{calrsfs}
\usepackage{tabularx}

\RequirePackage[abort, orthodox]{nag}

\usepackage{amsmath}
\usepackage{xcolor}
\usepackage{booktabs}
\usepackage{currfile}
\usepackage{enumitem}
\setlist{nosep,leftmargin=\parindent}
\usepackage{hhline}

\usepackage{expl3}
\usepackage{float}

\usepackage[binary-units, group-separator={,}]{siunitx}
\usepackage{subcaption}
\usepackage{tabularx}

\usepackage{wrapfig}

\usepackage{bookmark}

\pgfplotsset{
  compat=1.13,
  input steering scatter plot/.style={
    font=\small,
    grid=minor,
    legend pos=south east,
    log ticks with fixed point,
    width=.25\textwidth,
    xmax=300,
    xmin=.3,
    ylabel style={
      align=center,
    },
    ymax=300,
    ymin=.3,
  },
}

\pgfplotstableset{
  code string type/.style={
    assign cell content/.code={%
      \pgfkeyssetvalue{/pgfplots/table/@cell content}{\lstinline{##1}}
    },
    column type=l,
  },
  col sep=comma,
  every head row/.style={
    before row=\toprule,
    after row=\midrule,
  },
  every last row/.style={
    after row=\bottomrule,
  },
  incidence numeric type/.style={
    column type={S[table-format=#1]},
    multicolumn names=r,
    numeric as string type,
  },
}

\usetikzlibrary{
  arrows.meta,
  calc,
  chains,
  decorations.pathreplacing,
  fit,
  positioning,
  scopes,
  shapes.symbols,
  tikzmark,
}
\sethlcolor{pink}
\DeclareSIUnit[number-unit-product={}]{\percent}{\%}
\ExplSyntaxOn{}
\newcommand*{\scaled}[1]{\fp_eval:n{#1 / 10000}}
\newcommand*{\pcent}[3][round-mode=figures]{\SI[#1]{\fp_eval:n{100 * #2 / #3}}{\percent}}
\newcommand*{\pcentSum}[3][round-mode=figures]{\SI[#1]{\fp_eval:n{100 * #2 / (#2 + #3)}}{\percent}}
\ExplSyntaxOff{}

\newcommand{\inputSteeringScatterPlot}[5][]{%
  \begin{tikzpicture}
    \begin{loglogaxis} [input steering scatter plot, xlabel=#3, ylabel=#5, #1]
      \addplot+ [only marks, mark=x] table [x=#2, y=#4] {\inputSteeringResults} ;
      \addplot+ [mark=none, domain=.3:300] {x} ;
    \end{loglogaxis}
  \end{tikzpicture}
}

\DeclareMathAlphabet{\pazocal}{OMS}{zplm}{m}{n}

\definecolor{backward}{HTML}{1B9E77}
\definecolor{existing}{HTML}{D95F02}
\definecolor{forward}{HTML}{7570B3}

\newcommand{\zw}[1]{\textbf{\color{green}\emph{Zi: #1}}}

\newcommand{\Omit}[1]{}

\begin{document}

\title[TOFU]{TOFU: Target-Oriented FUzzer}         



\author{Zi Wang}
\email{zw@cs.wisc.edu}

\affiliation{%
  \institution{University of Wisconsin--Madison}
  \city{Madison}
  \state{Wisconsin}
}

\author{Ben Liblit}
\email{liblit@cs.wisc.edu}

\affiliation{%
  \institution{University of Wisconsin--Madison}
  \city{Madison}
  \state{Wisconsin}
}

\author{Thomas Reps}
\email{reps@cs.wisc.edu}
\orcid{0000-0002-5676-9949}

\affiliation{%
  \institution{University of Wisconsin--Madison}
  \city{Madison}
  \state{Wisconsin}
}


\begin{abstract}
Program fuzzing---providing randomly constructed inputs to a computer
program---has proved to be a powerful way to uncover bugs,
find security vulnerabilities, and generate test inputs that
increase code coverage.
In many applications, however, one is interested in a \emph{target-oriented}
approach---one wants to find an input that causes the program to reach
a specific target point in the program.
We have created TOFU (for \textbf{T}arget-\textbf{O}riented
\textbf{FU}zzer) to address the directed fuzzing problem.
TOFU's search is biased according to a distance metric that scores each input
according to how close the input's execution trace gets to the target locations.
TOFU is also \emph{input-structure aware} (i.e., the search makes use
of a specification of a superset of the program's allowed inputs).

Our experiments on \texttt{xmllint} show that TOFU is 28\% faster than AFLGo, while reaching 45\% more targets.
Moreover, both distance-guided search and exploitation of
knowledge of the input structure contribute significantly to TOFU's
performance.

\end{abstract}



\maketitle

\section{Introduction}%
\label{Se:Introduction}

Fuzz testing is an automated technique for testing a program by
generating random inputs and running the program on those inputs.
It is widely used for identifying bugs and security vulnerabilities,
and for generating test inputs that increase code coverage.

Even when used for identifying bugs and security vulnerabilities,
the focus of \emph{standard fuzz testing} is on increasing code coverage:
more code covered generally results in more bugs found.
However, in many applications one wants to find an input that causes
the program to reach a specific target point in the program (or set of target points).
For instance, one might be interested in (i) generating inputs that exercise
newly patched code, or (ii) as a way to reproduce a crash, generating inputs
that exercise regions of code known to be likely to cause a crash. 
In other words, one is interested in a directed approach,
in which the target points are given a higher priority
than other parts of the program.
The goal of a \emph{target-oriented fuzzer} is to spend the available
computational resources on the task of reaching those specific points
in the program.

\begin{focusbox}
  Given some target locations in the source code, craft a set of inputs
  that causes these locations to be reached during execution.
\end{focusbox}

Unfortunately, the techniques used in standard fuzzers offer
little control over which parts of the program are explored.
For example, American Fuzzy Lop (AFL)~\citep{afl} is a state-of-the-art
standard fuzzer.
However, AFL does not make use of any information about the 
structure of the program's inputs, and thus creates many inputs
that are rejected by the program.
This approach is less useful for finding inputs that reach
specific locations---particularly if the targets are
deep in the part of the program that constitutes its
core computation, rather than in the program's
input-validation code.

AFLGo~\citep{Bohme:2017:DGF:3133956.3134020} is a system
that supports target-oriented fuzzing.
However, many of AFLGo's features are inherited from the American Fuzzy Lop (AFL)~\citep{afl};
in particular, AFLGo does not use any information about the 
structure of the program's inputs.

Most modern fuzzers share the same structure:
a loop involving mutation, feedback, and evaluation of fitness.
AFL is an example:
at each step, the input is mutated;
if the mutant causes an execution run that has a new edge-coverage profile,
the mutant is retained as a seed for further mutation.

Fuzzing works well because small changes in inputs usually cause a small
change in the parts of the program that are exercised.
For target-oriented fuzzing, this ``stability heuristic'' can be used
to steer the fuzzer toward an input that reaches the desired target:
by focusing on inputs that are closer to the target, the fuzzer
might generate inputs that are even closer.
The hope is that this process will eventually allow the target to be reached.

Unfortunately, random changes to some bits or bytes of the input can
significantly change the program's execution;
for instance, such a change can turn a valid input into an invalid input
(leading to an execution run that is usually much farther away from
the desired target).

For target-oriented fuzzing, one way to restore the stability heuristic
is by taking advantage of known structure of the program's inputs:
the heuristic is now that small \emph{structural} changes in inputs
usually cause a small change in the parts of the program that are exercised,
and the fuzzer can take advantage of such behavior to steer itself to
an input that reaches the desired target.

\paragraph{TOFU\@: {\bf T}arget-{\bf O}riented {\bf FU}zzer.}
In our work, we have created a new directed fuzzer, called TOFU.
The high-level structure of TOFU is similar to that of AFL\@;
however, there are two important differences.
\begin{enumerate}
  \item
    TOFU's goal is
    to produce inputs that reach a specific set of targets in the program.
    TOFU pre-computes the distance between each pair of basic blocks,
    and uses these distances, together with the outcome of executing
    various inputs, to determine the order in which inputs are selected for mutation.
  \item
    TOFU leverages knowledge of the program's input structure, which is
    provided by the TOFU user in the form of a protobuf specification \citep{protobuf}.
\end{enumerate}

We are interested in understanding how much these two aspects of TOFU
contribute to its overall effectiveness, which gives rise to the following research questions:

\begin{rqbox}{guidance}
  What is the contribution of distance-guided search to TOFU's overall effectiveness?
\end{rqbox}

\begin{rqbox}{structure}
  What is the contribution of structured mutation to TOFU's overall effectiveness?
\end{rqbox}

The existing AFLGo and Hawkeye~\citep{Chen:2018:HTD:3243734.3243849} tools are also
target-oriented fuzzers.
Their high-level organization is similar to TOFU's in that the source code
is first analyzed to generate a static metric, and then the metric is used to guide the
subsequent process of dynamic exploration.
However, TOFU addresses the following issues in a different way than existing tools:

\begin{enumerate}
  \item
    The input space for both AFLGo and Hawkeye is a single file or
    \texttt{stdin}, as in AFL\@.
    However, the input space for many programs is larger
    than a single file.
    For instance, a program may use command-line flags to indicate option settings.
    AFL is designed to maximize coverage, which means that it might need
    different flag settings:
    different flag settings can lead to execution runs that cover different parts of the program.
    However, a directed fuzzer has a different goal, namely,
    to reach a specific target, or set of targets, in the program.

    \hspace*{1.5ex}
    To reach specific locations in the program, it may be necessary to
    provide specific flags on the command line.
    TOFU augments the input space that it explores to include command-line flags, so that
    users do not have to select such flags manually.
    (This aspect of TOFU's approach also reduces what the user has to understand about the program.)

  \item
    One wishes to generate inputs that reach desired targets
    as quickly as possible.
    AFLGo introduced scheduling via simulated annealing:
    i.e., inputs closer to the target get mutated more.
    Hawkeye adopted scheduling via simulated annealing and added prioritization:
    i.e., inputs closer to the target are mutated first.
    The Hawkeye work showed that, without prioritization,
    an input that nearly reaches a target might
    wait for a long time before getting mutated (cf.\ \citet{Chen:2018:HTD:3243734.3243849}).

    \hspace*{1.5ex}
    TOFU only applies prioritization, but uses a different metric
    than the ones used in AFLGo and Hawkeye.
    TOFU's metric has an intuitive interpretation as a distance (e.g., the number
    of correct branching decisions needed to reach the target),
    whereas scheduling in AFLGo (as well as prioritization in Hawkeye)
    has a complicated relationship to the history of the annealing that
    has taken place, and has no intrinsic meaning as a distance.

  \item
    Both AFLGo and Hawkeye use the mutators from AFL, which is
    general-purpose, but not structure-aware.
    The mutation operations can create many invalid inputs.
    This approach is useful for triggering buggy behavior
    \emph{somewhere} in the program (which is appropriate for
    standard fuzzing), but it is
    hard for a structure-blind fuzzing tool to find an input that passes
    the input-validation tests typically performed by a program,
    and even harder for it to find an input that reaches a desired
    target.

    \hspace*{1.5ex}    
    Because TOFU mutates inputs with respect to a specification of
    the input language, it has an easier time finding valid inputs,
    which (often) allows it to find inputs that reach a target that
    is deep in the program.
\end{enumerate}
These differences lead to the following research question:
\begin{rqbox}{existing}
  How does the performance of TOFU compare to existing tools?
\end{rqbox}
We found that the time taken by TOFU (pre-fuzzing) to compute distance information and to insert instrumentation was $\sfrac{1}{40}$ the time taken by AFLGo.
Moreover, the success rate for TOFU, measured by how many target basic blocks are covered,
is 45\% higher than that of AFLGo (Table~\ref{table:xmllint}). TOFU is also 28\% faster than AFLGo (Figure~\ref{fig:xmllint}).

\paragraph{Command-line Fuzzing and Staged Fuzzing.}
In response to the common practice of using command-line flags to
specify options to a program, TOFU includes command-line arguments in
the input space that it explores.
Prior fuzzing tools have not emphasized exploration of these inputs.
Command-line flags are commonly used by many programs,
while most fuzzing tools focus mainly on one input file
used by the program under test
(which is often specified by some file argument on the command line).
It is often necessary to supply the program with specific flags
for particular target basic blocks to be reachable.
A user could read the source code to try to understand what flags must
be set;
however, the targets could be any locations in the source code, and
the manual approach could require a big investment of time.
To address this issue, TOFU first performs a search---which is really
just a variant of its ``core search'' using a distance metric and
structured mutation---to find appropriate flag settings and options to flags
that get \emph{close} to the target basic blocks.
(In fact, in 60\% percent of the cases, command-line fuzzing
alone succeeded in reaching one or more of the target basic blocks.)

Our experiment on \texttt{libxml2}\citep{libxml2} showed the effectiveness of dividing the
fuzzing process into stages. In particular, fuzzing the flags and
fuzzing the primary input file can be carried out in sequence.
This approach reduces the dimensionality of the input space for
each individual stage of fuzzing, and we found that fuzzing efficiency
is improved by doing so.
The idea of staged fuzzing also appeared in Zest~\citep{Padhye:2019:SFZ:3293882.3330576},
in which the fuzzing process is divided into syntactic-fuzzing and semantic-fuzzing
stages.
We believe that the staged-fuzzing idea can be further exploited if
the input space can be divided into more fine-grained stages.


\paragraph{Contributions.}
Our work makes three main contributions:
\begin{enumerate}
  \item
    We use structured mutation to address the problem of target-oriented fuzzing.
  \item
    We expand the idea of \emph{staged fuzzing}.
    TOFU augments the search space for fuzzing to include
    a program's command-line flags.
    To reduce the dimensionality of the search carried out,
    command-line fuzzing is carried out separately, and prior to, the
    fuzzing of the program's primary input file.

    \hspace*{1.5ex}
    To support this task, TOFU provides a \emph{generator of structured mutators} for a
    family of command-line languages typical of those used by many Linux programs.
    TOFU---equipped with such a mutator---could be used
    as a pre-processing step to select appropriate command-line
    flags, before letting another fuzzing tool take over
    to fuzz the program's primary input file.
  \item
    The tool chain of TOFU consists of multiple components that can be reused
    in other tools.
    For example, TOFU's phase of distance computation takes only about $2$ minutes for \texttt{libxml2},
    while AFLGo's distance computation takes about $80$ minutes.
    Moreover, the distance computation is not a one-time investment---it is
    needed for each set of targets.

    \hspace*{1.5ex}
    TOFU also comes with specifications of different languages of structured
    inputs, which can be reused for fuzzing multiple application programs.
    For instance, the input language for \texttt{libxml2} is XML;
    thus, specification that we created for fuzzing \texttt{libxml2}
    can be reused for any application that takes an XML file as input.
\Omit{
    For example, in on-going work, the ``\texttt{space}'' mutator has
    been used to experiment with improving the robustness of a test suite:
    a mutation-testing tool is used (repeatedly) to create a mutant of \texttt{space} (i.e., $\texttt{space}'$);
    TOFU is used to attempt to kill $\texttt{space}'$ by finding an input $i$ with
    different observable behavior on \texttt{space} and $\texttt{space}'$;
    input $i$ is then added to the test suite.
}
\end{enumerate}

\paragraph{Organization.}
The remainder of the paper is organized as follows:
\sectref{Examples} presents two examples to motivate the problem that TOFU addresses.
\sectref{Overview} gives an overview of how instrumentation and execution is carried out in TOFU\@.
\sectref{DetailedDescription} presents the details of the algorithms used in TOFU\@.
\sectref{Experiments} presents experimental results.
\sectref{ThreatsToValidity} discusses threats to the validity of our results.
\sectref{Discussion} discusses different aspects of TOFU.
\sectref{Related-Work} discusses related works.
\sectref{Conclusion} concludes.



\section{Examples}
\label{Se:Examples}

\paragraph{Structured Mutator}
Consider the program shown in \figref{SMExample},
and suppose that its input language is described by the grammar
\begin{equation}
  \label{Eq:LinearGrammar}
  S  \rightarrow aSa \,|\, bSb \,|\, \epsilon
\end{equation}

\begin{figure}
  \centering
\begin{lstlisting}[language=C, numbers=left, xleftmargin=2em, escapechar=|, showstringspaces=false]
bool is_valid(string input){
# validate the input with respect to grammar |(\ref{Eq:LinearGrammar})|
}

int num_a(string input){
 # count how many number of a's in input
}

int main(int argc, char** argv){
  string input = argv[1];
  if(!is_valid(input))
    perror("Invalid input\n");|\label{Li:InvalidInput}|
    
    ...
  if(num_a(input) == 10){
    # target location   |\label{Li:TargetLocation}|
  }	
}
\end{lstlisting}
  \vspace{-1.5ex}
  \caption{\label{Fi:SMExample}Example for structured mutator.}
\end{figure}

\noindent
The program first validates the input;
then performs other computations;
and finally counts the number of occurrences of the character \texttt{a} in the input.

This example illustrates why a fuzzer can perform much better if it mutates the input
with respect to the input grammar.
For execution to reach the target location in line \ref{Li:TargetLocation},
the input must be grammatically correct---i.e., it must
be a string in $L(S)$ of grammar (\ref{Eq:LinearGrammar})---and contain exactly
10 occurrences of the character ``\texttt{a}''.
Any single-character mutation of a grammatically valid input creates
an invalid input that will be rejected by the program in line~\ref{Li:InvalidInput}.
Therefore, even if a valid input yields an execution that gets close
to the target location, a single-character mutation will create
an input that is never close to the target location.
Moreover, for a single-character mutation to create an input that
reaches the target location, it must start with an invalid input
that would be rejected by the program (and hence one whose execution
is never close to the target location).

\begin{figure}
  \centering
\begin{lstlisting}[language=C, numbers=left, xleftmargin=2em]
static char const shortopts[] =
  "0123456789abBcC:dD:eEfF:hHi" \
  "I:lL:nNpPqrsS:tTuU:vwW:x:X:y";

bool ignore_blank_lines = false;

int main (int argc, char **argv) {
  while ((c = getopt_long (argc, argv, shortopts, longopts, NULL)) != -1) {
    switch (c) {
      case 'B':
        ignore_blank_lines = true;
        break;
      ...
    }
    exit_status = compare_files (...);
  }
}

static int compare_files (...) {
  ...
  status = diff_2_files (&cmp);
  ...
}

int diff_2_files (struct comparison *cmp) {
  ...
  if (ignore_blank_lines) {
    if(analyze_hunk(...))
        #target location
    else ...
  }
  ...
}
\end{lstlisting}
  \vspace{-1.5ex}
  \caption{\label{Fi:CmdExample}Example for command-line fuzzing.}
\end{figure}

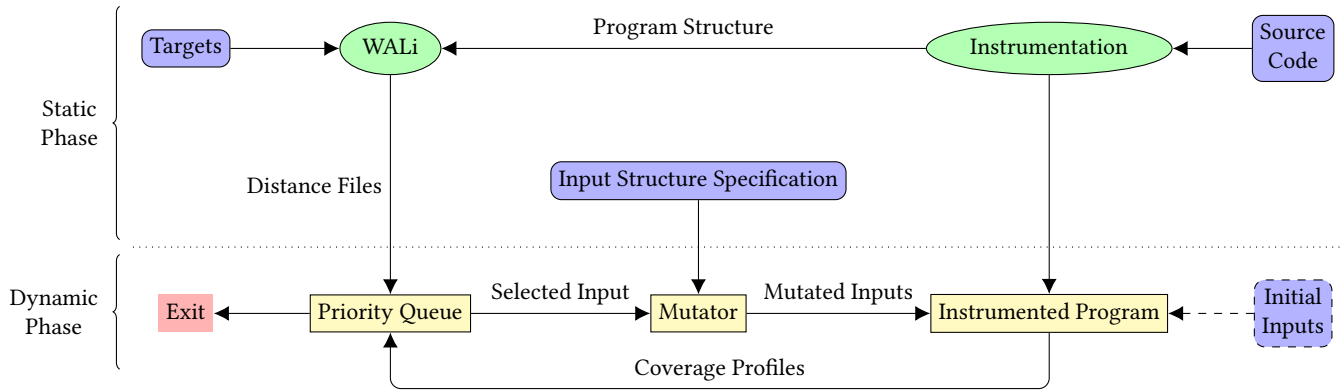
\begin{figure*}
  \centering
  \begin{tikzpicture}
    [
    align=center,
    >={Latex[width=2mm,length=2mm]},
    base1/.style = {
      align=flush center,
      draw=black,
      rectangle,
      rounded corners,
    },
    base2/.style = {
      align=flush center,
      draw=black,
      ellipse,
    },
    base3/.style = {
      align=flush center,
      draw=black,
      rectangle,
    },
    base4/.style = {
      align=flush center,
      draw=none,
    },
    input/.style = {base1, fill=blue!30},
    op_input/.style = {align=flush center,
      draw=black, dashed,
      rectangle,
      rounded corners, fill=blue!30},
    static/.style = {base2, fill=green!30},
    fuzzing/.style = {base3, fill=yellow!30},
    exit/.style = {base4, fill=red!30},
    ]

    \matrix [row sep=30, column sep=30] (grid) {
      \node [input] (tg) {Targets} ; &
      \node [static] (wali) {WALi\vphantom{y}} ; &
      &
      \node [static] (inst) {Instrumentation\vphantom{y}} ; &
      \node [input] (sc) {Source\\Code\vphantom{y}} ;
      \\
      & & \node [input] (iss) {Input Structure Specification} ;
      \\
      \node [exit] (ex) {Exit\vphantom{y}} ; &
      \node [fuzzing] (pq) {Priority Queue} ; &
      \node [fuzzing] (mt) {Mutator\vphantom{y}} ; &
      \node [fuzzing] (isp) {Instrumented Program} ; &
      \node [op_input] (in) {Initial\\Inputs} ;
      \\
    } ;

    \draw [->, rounded corners]
    (tg) edge (wali)
    (inst) edge node [above] {Program Structure} (wali)
    (sc) edge (inst)
    (wali) edge node [left] {Distance Files} (pq)
    (inst) edge (isp)
    (iss) edge (mt)
    (pq) edge (ex)
    (pq) edge node [auto] {Selected Input} (mt)
    (mt) edge node [auto] {Mutated Inputs} (isp)
    (in) edge [dashed] (isp)
    (isp) -- +(0, -1) -| node [above, near start] (profiles) {Coverage Profiles} (pq)
    ;

    \coordinate (strata) at ($(iss)!.5!(mt)$) ;
    \coordinate (strata west) at (grid.west |- strata) ;
    \coordinate (strata east) at (grid.east |- strata) ;
    \draw (strata west) edge [dotted] (strata east) ;

    \draw [decorate, decoration={brace, raise=5}] ($(strata west) + (0, .1)$) -- node [auto, xshift=-10] {Static\\Phase} (grid.north west) ;
    \draw [decorate, decoration={brace, raise=5}] (grid.west |- profiles) -- node [auto, xshift=-10] {Dynamic\\Phase} ($(strata west) - (0, .1)$) ;
    ;

  \end{tikzpicture}

\caption{\label{Fi:workflow}
  Workflow of TOFU:
  Round-cornered nodes indicate inputs for TOFU, where initial inputs are optional.
  Elliptical nodes denote elements of TOFU's static phase.
  Square nodes with a frame indicate TOFU's dynamic/fuzzing phase.
  The square node without a frame indicates the exit.
}
\end{figure*}

\paragraph{Staged Fuzzing}
We present a code snippet to illustrate the staged-fuzzing idea and
how our flag fuzzer works.
The code snippet in \figref{CmdExample} is a slightly modified
version of code found in the \texttt{diff} program in \texttt{diffutils}\citep{diffutils}.
The input for \texttt{diff} consists of command-line flags with options
and two files (or directories, or \texttt{stdin}).
Suppose that the target is in the \texttt{diff\_2\_files} function,
as shown in the code snippet;
to reach the target, it is necessary that \texttt{ignore\_blank\_lines} be set to
\texttt{true}, which requires \texttt{diff} be invoked with the flag \texttt{-B}.
Consider two executions of \texttt{diff} in which one contains the flag \texttt{-B}
in the input, and the other does not.
In the first case, execution enters the \texttt{if} condition at line 27, while
in the second case, execution does not enter the \texttt{if} condition.
Therefore, the basic-block coverage in the first case is closer
to the target location than the basic-block coverage in the second case.
By measuring the distance, TOFU can decide that, to reach the target,
\texttt{-B} is likely needed as part of the command line.


\section{Overview of TOFU}%
\label{Se:Overview}

\Cref{Fi:workflow} shows TOFU's overall workflow.
TOFU performs a \emph{static phase}, during which it (i) instruments the source code,
and (ii) pre-computes distances in a modified interprocedural
control-flow graph, and a \emph{dynamic phase}
during which it mutates inputs and executes the program.
As depicted in \Cref{Fi:workflow} via the round-cornered nodes, TOFU
takes as inputs the source code, the target lines, and the
input-structure specification.
During the static phase (above the dotted line), TOFU instruments the program so that at runtime,
it can extract the basic-block coverage from each execution run of the program.
In addition, TOFU generates distance files that record the pairwise distance from each basic block
in the instrumented program to each target basic block.
These files are used during the dynamic phase to guide input selection.
Moreover, because fuzzing during the dynamic phase is structure-aware,
TOFU also takes in a specification of the structure of allowed inputs,
and generates the structured mutator that is used in the dynamic
phase.

During the dynamic phase (below the dotted line), the fuzzer executes the instrumented program,
and assigns scores to the different inputs based on the executed basic blocks and the distance metric.
All of the inputs generated are stored in a priority queue, where the priority indicates
how close the trace obtained from the input came to the target set.
On each round, the closest input is selected from the priority queue for mutation;
the structured mutator then generates a user-specifiable number of new
mutated inputs.

The dashed frame of the node labeled ``Initial Inputs'' in \Cref{Fi:workflow} indicates that the
initial inputs are optional.
Because TOFU uses a structured mutator, it does not have to start with initial inputs;
instead, it can generate them according to the structural specification.
The fuzzing process performs the loop shown in the ``Dynamic Phase'' portion of \Cref{Fi:workflow}.
Fuzzing terminates when each target basic block is reached by some input,
or a timeout threshold is exceeded.



\section{Detailed Description}%
\label{Se:DetailedDescription}

\subsection{Gray-Box Fuzzing with Prioritization}
\label{Se:GrayBoxFuzzingWithPrioritization}
The fuzzing process is a repeated loop of mutation and execution, until all
target basic blocks are reached or the timeout threshold is exceeded.
During each iteration, TOFU selects a seed input and generates multiple new
mutated inputs based on the selected input.
The number of mutated inputs generated in each mutation round
can be defined by the user.
The mutated inputs are similar to the seed input, and thus the
coverage induced by each mutant is likely to be similar to the
coverage induced by the seed input.
The hope is that each time TOFU creates mutants from the input selected as the seed,
the induced coverage for \emph{some} of the newly mutated inputs will
get closer to the target set.
In this way, the program's execution traces are likely to get
gradually closer to---and eventually reach---members of the target
set.

TOFU uses a priority queue to store candidate inputs, and the
priority is a function of both the static distance files
generated in the static-analysis phase and the coverage induced by
each input.
At the beginning of the dynamic phase, TOFU reads the
pre-computed distance files generated during the static phase.
A distance file $M_t$ contains the distances from each basic block $b$
in the program to a target block $t$.
The distance $d_{b,t}$ represents the minimum number of choices that
would have to be made correctly, for an input that causes execution to
reach $b$, to continue on to $t$.

When a mutant $j$ is generated from a seed input $i$, TOFU executes
the instrumented program with $j$, and the coverage produced by the
program is retrieved by TOFU\@.
The fitness score for a pair $(j,t)$, where $t$ is some target block,
is the minimum of the distances induced by all of the basic blocks
reached during the execution of $j$.
The interpretation of this score is that it represents the minimum
number of additional choices that a mutation of $j$ would have to make
correctly to reach $t$.
The priority for $j$ is the minimum of $j$'s fitness scores
across all of the target blocks.

\paragraph{Distance Computation.}

The goal of the distance computation is to count how many constraints
remain to reach the target location for each basic block.

Given the inter-procedural control-flow graph (ICFG) of the
instrumented program, to compute such distances we label the graph's
edges with lengths---and introduce new edges---as follows:
if an edge has no branches, we give the edge the length $0$;
otherwise, we give the edge the length $1$.
(The edges involving branches are the control-flow edges
that have more than one successor, and the indirect-call edges
that have more than one potential callee.)
Additionally, we also add edges with length 0 from a basic block
to its immediate post-dominators.
We then compute the shortest distance in the modified ICFG
between each basic block and each target basic block.
(If block $b_2$ is not reachable from $b_1$, then the distance from $b_1$ to $b_2$
will be $\infty$.)

An ideal distance computation would be context-sensitive:
when computing the distance from an input $i$ to a target $t$:
only feasible paths from each basic block in the execution trace of $i$ to $t$ should be considered.
However, this approach would require TOFU to record the \textit{call stack} for each element of $i$'s execution trace.
Thus, to reduce the run-time overhead of instrumentation, only information
about executed \textit{basic blocks} is reported;
in particular, the contents of the call stack are \emph{not} reported
for each basic block encountered.
We use a heuristic to approximate the context-sensitive metric.
The intuition for the heuristic is that only functions that
can reach the target locations in the call graph are considered in the
distance computation.
Basic blocks in functions that cannot reach any target location
via some call chain are considered irrelevant.
Therefore, if there is no call path from \texttt{main} to the target locations
via a given call edge, we give the call edge the length $\infty$
in the modified ICFG.

In TOFU, the distances used for prioritization are computed by a
generalization of Dijkstra's algorithm \cite[\S6.5]{CSFW:SJRS03};
it computes interprocedural (context-sensitive) distances in the modified ICFG.

\subsection{Structured Mutation}

Mutation-based directed fuzzing relies on the heuristic that although mutating
an input generates a similar input, the coverage of the mutated input
can be different from the coverage of the original input.
Thus, selecting those mutants whose coverage is closer to one of the
target blocks is likely to produce an input that is closer to a desired
input (i.e., one that reaches a target block).
Iterating this process may eventually generate a desired input.

A structure-blind mutator treats the input as a stream of bytes,
modifying the bytes little-by-little, but it can also transform
a valid input into a syntactically invalid input.
Structured mutation modifies an input with respect to its underlying
structure.
Structured mutation has recently been used by AFLSmart~\citep{aflsmart}
and Superion~\citep{superion}.
However, the goal of structured mutation in those two tools is to
increase the coverage when using fuzzing to test programs.
TOFU uses structured mutation to support directed fuzzing.

\subsubsection{Protobuf-Based Structured Mutation}
\label{Se:ProtobufBasedStructuredMutation}

In TOFU, the structured mutator is based on the Google
\texttt{libprotobuf-mutator} \citep{google_libprotobuf-mutator_2019}.
Protocol buffers (``protobufs'') are a ``language-neutral,
platform-neutral extensible mechanism for serializing
structured data''~\citep{protobuf}.
Google also developed \texttt{libprotobuf-mutator}, a tool to randomly
mutate protobufs.

To use \texttt{libprotobuf-mutator} in the context of TOFU,
a TOFU user provides a
\emph{standard protobuf-specification file}
that describes the structure of the inputs to the program.
This specification file is then compiled into a C++ class $C$.
A program input corresponds to an object of class $C$;
the mutator generated via \texttt{libprotobuf-mutator} operates on
such objects:
it modifies a given object into a mutant object.

A user also provides a \emph{renderer function} that transforms an
object of class $C$ into an input (in the form of text).
This function renders the mutated object as the corresponding input text,
on which TOFU's dynamic phase can then execute the instrumented program.

If a user wishes to start TOFU with some initial inputs, then they
must provide a \emph{parser function} that transforms initial inputs
into the corresponding object of class $C$. 

To summarize, if the user is familiar with the input language, 
it usually takes one or two days to implement the input-language grammar
and the renderer function.
In the authors' experience, the most difficult part is the parser function.
In general, however, the program under test contains code that parses the
input, which can serve as the starting point for the parser function
for TOFU.
If the user decides not to use initial inputs, a parser function is not
required.
Once all the components of a structured mutator have been created,
they can be used with all programs that use the same input language.
This re-usability is an added bonus when one expends the effort to
specify a structured mutator for TOFU.

\subsubsection{Command-Line-Language Structured Mutator}

Command-line flags are commonly used in many programs.
In many AFL variants, users must specify these flags when fuzzing the
program.
Compared to standard fuzzing (i.e., fuzzing intended to increase coverage),
having appropriate flag values is more important in target-oriented fuzzing
because some target blocks can only be reached when certain flags are given.
For efficiency reasons, it is also desirable to use the minimal number
of flags and options, so that a fuzzer does not explore irrelevant
parts of the program and mutate unnecessary parts of the input.

It is inefficient to use a general-purpose mutator to mutate command-line input.
Fortunately, many command-line languages are both simple and highly
structured, using flag-names with dashes and taking options listed after the flag.
While creating a structured mutator for a command-line
language manually is not overly difficult, this task can be tedious and
error-prone.
To make the task easier, TOFU provides a tool that takes a
specification of the command-line flags and options, and
generates a structured mutator for the command-line language.
This mutator is used in a search---controlled in TOFU's standard way,
by the distance metric and structured mutation---to determine the flag settings and options that cause execution to get closest to the target basic block.

\begin{figure}
\begin{lstlisting} [numbers=left, xleftmargin=2em]
--silent|optional|no option
-v|optional|no option
--version|optional|no option
--help|optional|no option
file1|required|directory|PATH_TO_DIRECT
file2|required|directory|PATH_TO_DIRECT
\end{lstlisting}
  \caption{Example input specification for the generator of
    command-line-language mutators.\label{Fi:cmd_input}}
\end{figure}

\figref{cmd_input} is taken from the experiment with
the \texttt{cmp} program from \texttt{diffutils}.
It shows an example of the kind of input specification that a user provides
to create a command-line-language mutator, using TOFU's
command-line-language mutator generator.
In \figref{cmd_input}, \texttt{-\,-silent}, \texttt{-v}, \texttt{-\,-version}, and \texttt{-\,-help}
are all optional valid flags without options for \texttt{cmd};
\texttt{file1} and \texttt{file2} are required arguments, which take
values from a directory.
The directory contains the files that \texttt{cmp} compares.

\Omit{
\subsection{Staged Fuzzing}
If the input space can be divided into different stages, then we can
fuzz different stages in sequence so that the overall input space for
fuzzing is reduced: in the best case, it becomes additive rather than
multiplicative.
For example, suppose that to reach some target during directed fuzzing,
the input needs to satisfy both constraints $C_1$ and $C_2$;
however, $C_2$ might reveal itself only when $C_1$ is satisfied.
Suppose that $C_1$ is related to the $I_1$ part of the input;
$C_2$ is related to the $I_2$ part of the input;
and $I_1$ and $I_2$ are disjoint.
Then before finding the right value of $I_1$, any fuzzing effort expended on
$I_2$ is wasted, and mutating both $I_1$ and $I_2$ simultaneously is less
efficient.
In this case, it is more desirable to mutate $I_1$ first and then $I_2$.

The input space of TOFU consists of both the command-line settings for flags
(with flag options) and the input files, and these two components are disjoint.
In the discussion above, they correspond to $I_1$ and $I_2$, respectively.
TOFU fuzzes the flags first.
During the stage of command-line fuzzing, the goal of fuzzing is not to
reach the target \emph{per se}, but just to get close to the target basic blocks.

The staged-fuzzing idea can be generalized to other kinds of inputs
that are relatively independent, such as configuration files.
}

\subsection{Refinements of Prioritization}

TOFU supports two mechanisms beyond basic prioritization of inputs via its priority queue.
\begin{itemize}
  \item
    To increase the ``diversity'' of the kinds of inputs considered during its search,
    TOFU incorporates a mechanism for suppressing some inputs that ``look too much like''
    inputs already considered.
    This mechanism uses basic-block coverage as a rough measure of a program's execution:
    inputs with different traces may have the same basic-block coverage;
    the measure of similarity between inputs $i$ and $j$ is based on their executions
    having identical basic-block coverage.
    Thus, in addition to the priority queue, TOFU's search uses a dictionary $D$ of
    (basic-block set, count) pairs to record how many inputs that cover the same
    set of basic blocks have been added to the priority queue.
    When a new mutated input $i$ is produced, TOFU obtains its
    coverage set $c_i$---the set of basic blocks reached during the
    execution of the program on $i$.
    By querying $D$ with respect to $c_i$, TOFU obtains the count $n = D(c_i)$,
    and inserts $i$ in the priority queue with probability $\sfrac{1}{(n+1)}$.
    This mechanism decreases the probability that TOFU
    repeatedly works with similar inputs.
  \item
    When an input $i$ with score $s_i$ is selected from the priority
    queue, $i$ is also inserted back into the priority queue,
    with a new score of  $1.2 \times s_i$.
    In this way, TOFU is able to mutate input $i$ again in the future;
    $i$ has an artificially changed priority, but it would still be
    prioritized over an input whose distance from the target set is 20\%
    greater than that of $i$.
\end{itemize}



\section{Implementation \& Evaluation}%
\label{Se:Experiments}

\subsection{Implementation}%
\label{Se:Implementation}

TOFU's uses the Whole-Program-LLVM (WLLVM)~\citep{wllvm} tool to compile the
subject program on which fuzzing is to be applied to a single LLVM bitcode
module for analysis. Several subsidiary analyses are performed, and an
instrumented binary is produced that, during an execution of the (compiled)
subject program on a particular input, allows determining the set of basic
blocks that were reached during execution.

In addition to instrumentation, the main analyses performed serve to compute the
static pairwise distances from each basic block in the instrumented program to
each target basic block.
\begin{itemize}
  \item
    TOFU performs the following simple indirect-call analysis
    to identify a set of potential callees at each indirect-call site:
    the analysis assumes that the set of functions possibly
    called at a given indirect-call site $s$ consist of
    (i) each function $f$ whose address has been taken, for which (ii) the
    type of $f$ exactly matches the pattern of argument-type(s)
    and return-type that occurs at $s$.
  \item
    The distance computation described in \sectref{GrayBoxFuzzingWithPrioritization}
    is performed by 
    (i) extracting the ICFG from WLLVM\@;
    (ii) computing immediate-post-dominator information for each procedure's CFG\@;
    (iii) modifying the ICFG as described in \sectref{GrayBoxFuzzingWithPrioritization}
    and labeling its edges with 0 or 1, as appropriate;
    (iv) encoding the modified ICFG as a weighted pushdown system, using
    the WALi-OpenNWA~\citep{walidev_wali_2019} library (the encoding is performed
    via a well-known technique \citep[\S2.2]{FSTTCS:RLK07});
    (v) computing interprocedural (context-sensitive) distances using a technique
    described in \citet[\S6.5]{CSFW:SJRS03}.
\end{itemize}

\subsection{Evaluation}
We aim to answer the research questions posed in \sectref{Introduction}, specifically:
\begin{enumerate}
    \item What is the contribution of distance guidance to TOFU's overall effectiveness?
    \item What is the contribution of structured mutation to TOFU's overall effectiveness?
    \item How does the performance of TOFU compare to existing tools?
\end{enumerate}

The goal of directed fuzzing is to cover all the target basic blocks
as quickly as possible.
We evaluated different tools by checking how many target basic blocks
were reached within a timeout limit.
For each run of each tool, we also determined how long it took to
reach each of of the reached targets.

The tested subjects are \texttt{space}~\cite{space_source},
\texttt{diffutils}~\cite{diffutils}, and \texttt{xmllint} from
\texttt{libxml2}~\cite{libxml2}.
\texttt{Space} is a well-established subject for software-engineering research,
and has been studied in several prior papers~\cite{FSE:WCM, ICSE:mutation}.
\texttt{Diffutils} and \texttt{libxml2} were used in the experiments
reported in the AFLGo paper~\citep{Bohme:2017:DGF:3133956.3134020}.

To answer research question (1), we created a variant of TOFU, called
UG-TOFU (for ``Un-Guided TOFU''), which assigns random numbers as
scores in the priority queue, instead of using a score based on
distance.
To answer research question~(2), we created another variant of TOFU, called
US-TOFU (for ``Un-Structured TOFU''), which uses a ``dumbed-down'' mutator
extracted from AFL,\footnote{AFL's mutator
  consists of three parts, which perform the following steps:
  (i) deterministic mutation steps,
  (ii) repeated steps of havoc/random mutations, and
  (iii) steps that splice two random inputs together at
  some random midpoint.
  The ``dumbed-down'' mutator uses a single havoc step each time it is invoked.
}
in place of TOFU's structured mutators (which are described in
\cref{Se:ProtobufBasedStructuredMutation}).
To answer research question (3), we ran TOFU, AFLGo, and Superion---an
existing fuzzer with a structured mutator for XML files---on
\texttt{libxml2}.\footnote{
  Hawkeye was not available to us due to legal restrictions.
}

Each experiment had same two-phase structure.  In phase 1, we ran TOFU to obtain
a set of command-line flags and flag options.
In phase 2, the command-line information obtained from phase 1 was used for all of the
tools in the experiment;
during phase 2, the subject program was fuzzed by each of the tools---i.e.,
each tool was applied to the subject program so that it mutated the program's
primary input file. \texttt{space} only uses a single file as input.
  Consequently, the experiment with \texttt{space} had no phase 1: it consisted only of phase 2, using
  the different tools.

Prior to phase 1 of each experiment, we used TOFU's generator of
command-line-language mutators to create an appropriate
command-line-language mutator for the subject program.
If there were options available from the program's test suite, then we
used those options as part of the specification provided to TOFU's
generator of command-line-language mutators.
Otherwise, we manually constructed a specification of the subject
program's command-line language by consulting the program's documentation.
During phase 1 fuzzing, TOFU was started from the empty initial input
(no flags or options).
The names of all input files from the subject program's test suite
are provided to the phase 1 fuzzer as potential elements
of the command line that the fuzzer is searching for.

For phase 2 fuzzing, we ran each tool three times to accommodate
variance in running times.
We call each run of all the tools a \emph{trial}.
For each trial, we randomly selected 10 inputs from the
subject program's test suite, and used that set as the initial inputs
for each tool for that trial.
That is, during a given trial, all tools were started with the same
set of initial inputs, flags, and flag options;
for a given tool, each trial starts with a different selection of
initial inputs.

For a given tool and a given trial, if a target basic block $t$ was
reached by the tool at least once, we say that the tool \emph{covered} $t$
in that trial;
if, for a given commit $c$, at least one new target block was covered, we say
that the tool \emph{covered} $c$ in that trial.

\subsection{Experimental Setup}
We ran the experiments on a workstation with twenty-four
Intel\textsuperscript{\textregistered} Xeon\textsuperscript{\textregistered} 
X5675 CPUs running at \SI{3.07}{\giga\hertz} and \SI{189}{\giga\byte} of memory.
All tools in these experiments are parallelizable, and each was given
full use of all CPU cores while running.

For the experiment with \texttt{space} (phase 2 only), and
\texttt{diffutils} (phases 1 and 2), we limit each
phase of each trial to 150 seconds.
For the experiment with \texttt{xmllint} (phases 1 and 2), we set the
timeout limit for each phase of each trial to 300 seconds.
To fully use all the available cores, the number of mutated inputs
produced in each round is set to 120 (which is a multiple of the
number of cores).

\subsubsection{Space}
The \texttt{space} program is from the Software-artifact Infrastructure
Repository~\cite{sir}.
\texttt{Space} consists of \num{9564} lines of C code (\num{6218} executable).
It implements an interpreter for an array-definition language (ADL).
It has a total of 39 versions.
One of the versions is the reference version, and each remaining
version contains a single fault.
We used each version's fault location as that version's target,
and hence this set of examples serves as a proxy for
the task of reproducing a crash.
Most of the faults correspond to less than three target basic blocks.
\Cref{table:space} summarizes the results.

\begin{table}
\caption{Experimental results for \texttt{space}\label{table:space}} 
\centering
\begin{tabular}{l*{3}{S[table-format=3]}}
  \toprule
  & \multicolumn{3}{c}{Trial} \\
  \cmidrule{2-4}
  Quantity & 1 & 2 & 3 \\
  \midrule
  Total target basic blocks & 119 & 119 & 119\\
  Target blocks covered by initial inputs & 51 &  48 & 50\\
  Uncovered target basic blocks &   68 & 71 & 69  \\
  Commits not totally covered &   17 & 15 & 16  \\
  \addlinespace
  \rowcolor{yellow}
  Basic blocks covered by TOFU &   21 & 65 & 63 \\
  Basic blocks covered by AFLGo  &  5 & 6 & 9 \\
  Basic blocks covered by UG-TOFU  &  2 & 0 & 18\\
  Basic blocks covered by US-TOFU   &  0 & 0 & 0 \\
  \bottomrule
\end{tabular}
\end{table}

\Omit{
\pgfplotstableread{space_times.csv}{\inputSteeringResults}

\begin{figure*}
  \inputSteeringScatterPlot%
  {tofu}{TOFU}%
  {aflgo}{AFLGO}%
  \hfill
  \inputSteeringScatterPlot%
  {tofu}{TOFU}%
  {ug-tofu}{UG-TOFU}%
  \hfill
  \inputSteeringScatterPlot
  {tofu}{TOFU}%
  {us-tofu}{US-TOFU}
  \caption{Comparisons of times, in seconds, for each tool when
    $\ge71\%$ of the total targets for each \texttt{space} commit are
    reached.  Each blue ``\textcolor{blue}{\texttimes}'' represents
    one pair of times.  The diagonal red ``\textcolor{red}{$y = x$}''
    line shows where equal execution times would appear.  Thus, each
    \textcolor{blue}{\texttimes} above the red line is a faster result
    for TOFU\@; each \textcolor{blue}{\texttimes} below the red line is
    a faster result for that plot's other tool.\label{fig:space}}
\end{figure*}
}

\subsubsection{diffutils}
We used the GNU \texttt{diffutils} as the subject program.
It contains four executables:
\texttt{cmp}, \texttt{diff}, \texttt{diff3}, and \texttt{sdiff},
which are all related to finding differences between files.
We choose all the commits from November 2009 to May 2012, and
for the $n+1^{\textit{st}}$ commit, the target blocks were all
new code introduced subsequent to the $n^{\textit{th}}$ commit.
The total number of target basic blocks is 333.
They are in the files util.c, diff.c, io.c, diff3.c, cmp.c, dir.c, context.c,
and analyze.c.
For the targets in diff.c, diff3.c, and cmp.c, most are reachable by
phase 1 fuzzing alone (i.e., by an appropriate choice of flags and options).
The files with the most target basic blocks are diff.c, io.c and dir.c,
each of which has more than 40 target basic blocks.
For each code patch, the target basic blocks are in general close to
each other---i.e., the distances between them are small
according to the metric used by TOFU.

The \texttt{diffutils} executables that we tested were \texttt{cmp},
\texttt{diff}, \texttt{diff3}.\footnote{
  We excluded \texttt{sdiff} for technical reasons.
  \texttt{sdiff} is really just a launcher for \texttt{diff}. 
  Rather than exiting normally or crashing, \texttt{sdiff}
   ``terminates'' by calling \texttt{execvp} to replace itself 
  with \texttt{diff}. Our coverage-recording instrumentation 
  does not trigger when a process self-replaces in this manner.
}
There are a total of
8 commits for \texttt{diff3},
5 commits for \texttt{cmp}, and
28 commits for \texttt{diff}.
During phase 1 fuzzing, all targets for \texttt{diff3} were reached;
all targets but one for \texttt{cmp} were reached; and
all targets from 14 of the 28 commits for \texttt{diff} were reached.

There are 14 commits for \texttt{diff} in which not all target basic blocks were
completely reached during phase 1 fuzzing.\footnote{
  The one remaining target of \texttt{cmp} can only be reached when the command-line
  input contains an error, which lies outside the capabilities of TOFU's command-line fuzzer.
  Thus, we excluded this example from this experiment.
} We ran TOFU, AFLGo, UG-TOFU and US-TOFU over the remaining 14 commits, and the 
result is summarized in\cref{table:diff}.

\begin{table}
  \caption{Experimental results for \texttt{diff}\label{table:diff}} 
  \centering
  \begin{tabular}{l*{3}{S[table-format=3]}}
    \toprule
    & \multicolumn{3}{c}{Trial} \\
    \cmidrule{2-4}
    Quantity & 1 & 2 & 3 \\
    \midrule
    Total target basic blocks & 220 & 220 & 220\\
    Target blocks covered by initial inputs & 71 &  71 & 71\\
    Uncovered target basic blocks &   149 & 149 & 149  \\
    Commits not totally covered &   14 & 14 & 14  \\
    \addlinespace
    \rowcolor{yellow}
    Basic blocks covered by TOFU &  57  & 66 & 59 \\
    Basic blocks covered by AFLGo  &  24 & 24 & 24 \\
    Basic blocks covered by UG-TOFU  &  40 & 40 & 38\\
    Basic blocks covered by US-TOFU   &  24 & 24 & 24 \\
    \bottomrule
  \end{tabular}
  \end{table}

\Omit{
  \pgfplotstableread{diff_times.csv}{\inputSteeringResults}

  \begin{figure*}
    \inputSteeringScatterPlot%
    {tofu}{TOFU}%
    {aflgo}{AFLGO}%
    \hfill
    \inputSteeringScatterPlot%
    {tofu}{TOFU}%
    {ug-tofu}{UG-TOFU}%
    \hfill
    \inputSteeringScatterPlot
    {tofu}{TOFU}%
    {us-tofu}{US-TOFU}
    \caption{Comparisons of times, in seconds, for each tool when
      $\ge40\%$ of the total targets for each \texttt{space} commit are
      reached.  Each blue ``\textcolor{blue}{\texttimes}'' represents
      one pair of times.  The diagonal red ``\textcolor{red}{$y = x$}''
      line shows where equal execution times would appear.  Thus, each
      \textcolor{blue}{\texttimes} above the red line is a faster result
      for TOFU\@; each \textcolor{blue}{\texttimes} below the red line is
      a faster result for that plot's other tool.\label{fig:diff}}
  \end{figure*}
}

\subsubsection{xmllint}
\texttt{Libxml2} is a project and library for working with XML, and
\texttt{xmllint} is a program from \texttt{libxml2} that can be used to parse
and validate XML files. We chose all commits from May 2012 to October 2014, and
for the $n+1^{\textit{st}}$ commit, the target blocks were all new code
introduced subsequent to the $n^{\textit{th}}$ commit. There are a total of 382
commits, of which 186 are \texttt{xmllint} code changes; TOFU's distance
computation indicated that the new code introduced in 180 of the commits are
apparently reachable.
The target basic blocks are in 38 files.
The files with the most target basic blocks are parser.c, xpath.c,
uri.c, encoding.c, tree.c, xmlschemastypes.c, and buf.c, each of which
has more than 200 target basic blocks.
Similar to \texttt{diffutils},
basic blocks in the same code patch are generally close to each other.

During phase 1 fuzzing, all target blocks from 53 commits
were completely reached,
leaving 127 commits containing targets for phase 2.
We ran TOFU, AFLGo, Superion, UG-TOFU, and US-TOFU on
these commits;
the results are summarized in \cref{table:xmllint}.\footnote{
  Superion only accepts initial inputs of size less than 10KB\@.
  Therefore, when running Superion, we discarded any randomly
  selected initial inputs that exceeded 10KB\@.
}

\begin{table}
\caption{Experimental results for \texttt{xmllint}\label{table:xmllint}}
\centering
\begin{tabular}{@{\hspace{0ex}}r@{\hspace{1.25ex}}l*{3}{S[table-format=4]}@{\hspace{0ex}}}
  \toprule
  && \multicolumn{3}{c}{Trial} \\
  \cmidrule{3-5}
   & Quantity & 1 & 2 & 3 \\
  \midrule
  1. & Total target basic blocks & 3064 & 3064 & 3064\\
  2. & Target blocks covered by initial inputs & 720 &  739 & 772\\
  3. & Uncovered target basic blocks & 2344   & 2325 & 2292  \\
  4. & Commits not totally covered &   127 & 127 & 127 \\
  \addlinespace
  \rowcolor{yellow}
  5. & Basic blocks covered by TOFU &   261 & 270 & 198 \\
  \rowcolor{yellow}
  6. & Extra commits covered by TOFU &   49 & 50 & 42 \\
  7. & Basic blocks covered by AFLGo  &  160 & 189 & 153 \\
  8. & Extra commits covered by AFLGo &   36 & 37 & 36 \\
  9. & Basic blocks covered by Superion & 161 & 140  & 98 \\
  10. & Extra commits covered by Superion &   32 & 29 & 24 \\
  11. & Basic blocks covered by UG-TOFU  &  202 & 214 & 178\\
  12. & Extra commits covered by UG-TOFU &   43 & 44 & 39 \\
  13. & Basic blocks covered by US-TOFU   &  159 & 143 & 114 \\
  14. & Extra commits covered by US-TOFU &   33 & 32 & 25 \\
  \bottomrule
\end{tabular}
\end{table}

\pgfplotstableread{xmllint_times.csv}{\inputSteeringResults}

\begin{figure*}
  \inputSteeringScatterPlot%
  {tofu}{TOFU}%
  {aflgo}{AFLGO}%
  \hfill
  \inputSteeringScatterPlot
  {tofu}{TOFU}%
  {superion}{Superion}
  \hfill
  \inputSteeringScatterPlot%
  {tofu}{TOFU}%
  {ug-tofu}{UG-TOFU}%
  \hfill
  \inputSteeringScatterPlot
  {tofu}{TOFU}%
  {us-tofu}{US-TOFU}
  \caption{Time comparisons (in seconds), for each tool when
    $\ge21\%$ of the uncovered targets for each \texttt{xmllint} commit are
    reached.  Each blue ``\textcolor{blue}{\texttimes}'' represents
    one pair of times.  The diagonal red ``\textcolor{red}{$y = x$}''
    line shows where equal execution times would appear.  Thus, each
    \textcolor{blue}{\texttimes} above the red line is a faster result
    for TOFU\@; each \textcolor{blue}{\texttimes} below the red line is
    a faster result for that plot's other tool.\label{fig:xmllint}}
\end{figure*}

We also carried out a more detailed analysis of the data, with the goal 
of evaluating TOFU's performance against each of the other tools,
all as phase-2 fuzzers.
The results are shown in\Omit{ \cref{fig:space}, \cref{fig:diff}, and}
\cref{fig:xmllint}.
For most commits, it is difficult for any tool to reach all target basic blocks,
so we needed a way to understand performance when not all targets are reached.
Moreover, we wanted to compare performance on commits that represented a fuzzing
challenge of ``appropriate difficulty.''
We define ``appropriate difficulty'' in terms of a \emph{coverage threshold}:
for each tool and each commit $c$, we gathered fine-grained
information about how many of the set $T_c$ of target basic blocks of
$c$ that were \emph{not} covered (by TOFU) during phase 1 \emph{were}
covered by the tool during phase 2.
\Cref{fig:xmllint} shows tool-against-tool plots
for the times to reach $\ge 21\%$ of the targets in $T_c$.

We could have used a coverage threshold different from $21\%$;
however, if the threshold is too high, then the study would just show
that each tool rarely reaches the threshold within the timeout limit.
If the threshold is too low, we only get information about
the time needed to reach a single target, which only provides
information about well the tools perform on ``low-hanging fruit.''

We picked $21\%$ by reasoning as follows:
as shown in lines 6, 8, 10, 12, and 14 of \cref{table:xmllint},
the tools covered different sets of commits that had not already
been completely covered during phase 1;
in trial 1, the five tools covered 53 extra commits in total,
covering 359 of the $1,689$ targets in those commits;
in trial 2, they covered 54 extra commits,
covering 360 of the $1,677$ targets;
in trial 3, they covered 49 extra commits,
covering 282 of the $1,491$ targets.
Thus, we set the coverage threshold to
$(359 + 360 + 282)/(1689 + 1677 + 1491) \approx 21\%$
This methodology---where the threshold was chosen based
on ``performance of the group in aggregate''---ensured that the
coverage threshold is challenging for each of the tools.

The results of the study are shown in \cref{fig:xmllint}. Each
``\textcolor{blue}{\texttimes}'' in a comparison plot in \cref{fig:xmllint}
shows information about the times used by two tools, $A$ and $B$, to reach
$\>21\%$ of the targets for some commit. There are a few additional subtleties
in the way this data was gathered.

\emph{Granularity:}
    Commits vary in their number of targets.
    For instance, if some commit has eight targets, we plot the time
    required for tool $A$ to reach any two targets against the time
    required for tool $B$ to reach any two (which are
    not necessarily the same two).

\emph{Ordering:}
    As mentioned earlier, we ran each tool three times on each commit,
    using three different sets of initial inputs.
    For a given commit, each tool was provided with the
    same three input sets.

    However, because the coverage threshold was set at a
    non-trivial percentage, many runs of the different tools
    lead to timeouts.
    When one tool's time is below the timeout threshold, and
    the other tool times out, a natural choice would be to plot
    the pair on the north or east edge of the comparison plot,
    as appropriate.

    We chose a somewhat different strategy, based on the observation
    that because each of the tools implements a different search
    strategy for mutating inputs (including TOFU, UG-TOFU, and
    US-TOFU), their searches diverge due to the different strategies,
    and so there is relatively little importance to requiring that the
    tools' runs be matched by their input set.  For this reason, we
    compare two tools' behaviors with respect to achieving the
    coverage threshold for a commit based on their \emph{ordered
      times}.  That is, when comparing tool $A$ with tool $B$, we
    create one ``\textcolor{blue}{\texttimes}'' for $A$'s fastest
    trial versus $B$'s fastest trial, one
    ``\textcolor{blue}{\texttimes}'' for $A$'s median-speed trial
    versus $B$'s median-speed trial, and one
    ``\textcolor{blue}{\texttimes}'' for $A$'s slowest trial versus
    $B$'s slowest trial.

\Omit{
Each tool and version produces up to
three times (which depend on the initial inputs, because some selection of
initial inputs could reach all targets for a version but another selection
does not. If we have fewer than three times to compare, we will use
the same method to compare them as described below). We give pairwise
comparisons between two tools. However, because there are randomness
involved in these tools, we ran each single experiment, and compare
the results by comparing the ordered times. Suppose that for a version
$v$, tool A produced time $t_1, t_2, t_3$ with $t_2 < t_1 < t_3$, and
tool B produced time $t'_1, t'_2, t'_3$ with $t'_3 < t'_2 <
t'_1$. Then when compare the performances of A and B on version $v$,
we pair the results as $(t_2, t'_3), (t_1, t'_2), (t_3, t'_1)$ and
draw the scatter plot.
\zw{The sets of initial inputs for each trial are different. Suppose for a version $v$ at trial 1, initial inputs cover all targets, while for trials 2 and 3, initial inputs do not. Because all tools in the same trial shares the same initial inputs, then tool A produced times $t_2, t_3$, and tool B produced times $t'_1, t'2$. We compare the pairs in the same way as we do when we have three times. }
}

\emph{Targets not reached:}
We also investigated why the targets of some commits were not reached.
A large portion of them are not reached because they are not related
to the primary input file.
For example, patched code from \texttt{xmlschemas.c} needs specific
XML schema files for the targets to be reachable.
Some of the code patches are identified by the distance computation
as being reachable from the entry of \texttt{xmllint}, but the conditions
for reaching them are never satisfied.
For example, some patched code is in a procedure that has a formal parameter whose value must be non-\texttt{NULL} for the code to be reachable.
However, as called from \texttt{xmllint}, the actual parameter is always \texttt{NULL}.
This situation arises because the patched code resides in a general-purpose
library, but \texttt{xmllint} does not exercise that library in its
full generality.

\subsection{Findings}

For \textbf{\cref{rq:guidance}}, the experiments show that distance guidance 
can improve TOFU's ability to reach
more targets. The comparison between TOFU and UG-TOFU in \cref{fig:xmllint}
shows that even without distance guidance, UG-TOFU is able to generate inputs to
reach the coverage threshold in many commits; however, the performance of
UG-TOFU is worse than that of TOFU\@.

As noted earlier, the locations of code changes in a commit are generally
close to each other;
thus, an input that reaches one target is likely to be similar
to an input that reaches others,
so mutations of a successful input may allow a fuzzer to reach other targets.
TOFU prioritizes inputs according to the number of correct branch choices
needed to reach some remaining target;
in contrast, UG-TOFU does use such distance information, and thus may not
explore these ``close-to-successful'' inputs immediately.

For \textbf{\cref{rq:structure}}, the experiments reveal that the structured-mutator component is extremely
important to TOFU's performance.
In all experiments, TOFU reaches more targets than US-TOFU, especially for \texttt{space},
where US-TOFU did not reach a new target basic block in any of the three trials.
It is also noteworthy that in experiments from both \texttt{space} and
\texttt{xmllint}, AFLGo's performance is better than US-TOFU's.
One possible reason is that US-TOFU's mutator does not implement all
of AFL's mutator:
US-TOFU's mutator only performs AFL's random-mutation step.
These results indicate that if the user does not want to go to the trouble of providing a
specification of valid inputs, then it is likely to be better to use AFLGo
instead of US-TOFU\@. (Conversely, it would be interesting to see whether AFLGo
would achieve any performance gain if it adopted or incorporated TOFU's distance
guidance.)

For \textbf{\cref{rq:existing}}, our experiments show that TOFU outperforms
existing tools. For the \texttt{xmllint} coverage threshold experiment, TOFU is
28\% faster than AFLGo, and 60\% percent faster than Superion, computed as the
geometric mean of the ratio of each pair. (All timeouts were counted as 300
seconds.) AFLGo does not use prioritization or a structured mutator. Superion is
a general-purpose fuzzing tool, aiming to improve coverage, rather than reaching
specific locations in the program.
For \texttt{diff}, TOFU's mutator includes a filesystem language, allowing TOFU
to create new directories and files, which is beyond the capabilities of
AFLGo.

On \texttt{space}, TOFU performs much better than AFLGo, while on \texttt{xmllint},
TOFU's advantage is not as dramatic.
To understand this difference, we examined \texttt{xmllint}'s source code,
and looked for differences between the targets covered by TOFU vs.\ AFLGo.
Many of the targets are parsing-related:
parser.c has the most target basic blocks among all source files.
AFLGo is good at reaching those targets.
In contrast, for targets that are not in the parsing
stage---for example, targets in \texttt{xinclude.c}---if the initial inputs do
not contain features relevant for reaching the targets, then AFLGo cannot reach
them, whereas TOFU is able to construct inputs that allow the targets to be
reached. Thus, whether TOFU is a better choice over AFLGo depends on the target:
if the target is shallow, and wild or erroneous input is required to reach the
target, then AFLGo is better; if the target is deep in the program, and requires
features that are not represented in the initial inputs, TOFU is a better
choice. In fact, the degree to which the data is scattered in the plot of TOFU
vs.\ AFLGo in \cref{fig:xmllint} suggests that TOFU and AFLGo have different
strengths when it comes reaching different kinds of targets.
As suggested in the Superion paper~\citep{superion}, another reason may be that
XML is weakly-structured, whereas the input to \texttt{space} is highly structured.
As a point of comparison, our protobuf specification for XML has 142 lines,
while our protobuf specification for the input language of \texttt{space} has 300 lines.
Thus, it may be that TOFU performs better on \texttt{space} than on \texttt{xmllint}
(relative to the performance of AFLGo) because\Omit{ creating an appropriate mutated
input for a more structured input language is easier than for an input language
with less structure.
Conversely,} it is more difficult for AFLGo to create the more structured inputs
needed to fuzz \texttt{space} (cf.\ the first example in \Cref{Se:Examples})\Omit{,
whereas the challenges for AFLGo and TOFU are more similar when creating
inputs for \texttt{xmllint}}.


\section{Threats to Validity}%
\label{Se:ThreatsToValidity}

Our empirical evaluation tries to adhere to practices recommended by
\citet{Klees:2018:EFT:3243734.3243804}.  For example, we evaluated each tool
multiple times with randomly selected seed inputs to reduce bias due to any
particular selection of inputs.
\Citet{Klees:2018:EFT:3243734.3243804} recommend running AFL-like
fuzzers for at least 24 hours per trial, but that would take half a
year for the 180 \texttt{xmllint} commits we used.  However, we
are not trying to maximize coverage as in general fuzzing.
TOFU may be useful for users with limited time, e.g., who want to
create coverage tests for code changes.

To compare the performance of TOFU and AFLGo with a longer search time,
we selected the ten versions from \texttt{space} with the most uncovered targets.
Instead of running for just 150 seconds, we ran the tools for 1500 seconds.
The result is that TOFU still reaches 80\% more targets than AFLGo
(compared with 4x--6x more targets for the 150-second limit---see \cref{table:space}).
\Omit{TOFU reaches most targets within 100 seconds, while AFLGo reaches most of the targets after 200 seconds.}

TOFU only selects targets that its distance computation indicates are reachable.
Our reachability analysis makes a few assumptions that may not hold for all C
programs.  For instance, TOFU uses function type-signatures to approximate the
callable set at each indirect-call site.  TOFU does not consider casts, which
could allow a differently typed function to be called.  However, the effect of
this imprecision is small.  For example, there were only 6 of 186 xmllint code-change 
commits for which TOFU considered all targets to be unreachable.
TOFU's indirect-call analysis is still an advance over AFLGo, which
performs no indirect-call analysis at all.

\Omit{
\zw{We test over three subjects because understanding the grammar of those subjects 
and implementing the structured mutator takes a certain amount of manual effort. 
For example, Superion~\citep{superion} implements only two structured mutators.
However, as pointed out earlier, for the same format of inputs, the structured mutator can be reused. 
In fact, our ongoing project also shows that the TOFU has a great potential to 
improve testsuite robustness, by generating inputs to differentiate programs
that are modified under mutation testing--source code mutated under a systematic
mutation. Furthermore, both \citet{superion} and \citet{aflsmart} point out
that the manual effort for constructing structured mutator can significantly 
boost efficiency.}
}


\section{Discussion}%
\label{Se:Discussion}

\emph{Structured Fuzzing:}
Our structured mutator is based on protobuf specifications, which cannot capture
all constraints on valid inputs.  We therefore augmented these rough specifications
with additional semantic information. For example, for \texttt{space} inputs,
one data field contains a number $k$ followed by exactly $k$ \texttt{PORT}
definitions.  Our protobuf approximation of this format does not have a separate
integer data field corresponding to $k$.  Instead, we count \texttt{PORT}
definitions and reconstruct $k$ accordingly when rendering a protobuf instance
as a program input.  Furthermore, we do not have a full specified set of
features for every aspect of the input.  For example, for \texttt{xmllint} inputs,
we do not have a full XML Pointer (XPointer) Language.
Instead, we manually construct some XPointers and map each XPointer into an
integer option for the XPointer sub-language.  One advantage of this approach
is that the user might have prior knowledge on what parts in the input are
essential to reach certain targets.  A restricted implementation can make the
fuzzer reach the targets faster.  Furthermore, protobufs allow inheritance, so
the user can specify different parts of the input separately, then either reuse
or extend specification fragments as needed.

\emph{Fuzzing as a Proxy for Symbolic Execution:}
In theory, symbolic execution and constraint solving could derive an input to
reach a program location.  However, it is always difficult for this approach to
scale well.  TOFU offers fuzzing as a proxy for symbolic execution.  TOFU's
distance metric effectively counts how many constraints must be solved to reach
some target.  But instead of solving those constraints, TOFU uses structured
mutation to generate input variants that \emph{may} have fewer constraints to
solve.

\emph{Choice of Distance Metric:}
The distance metric we used counts how many branches remain to reach
the target location.
We also ran our experiments using an alternative metric in which
distance is the shortest distance in the ICFG from
a given basic block to the target basic block.
The results were that for most targets, the performance was nearly the same,
and the differences observed did not reveal that one metric
is preferred over the other.
\Omit{
This is because we use a priority queue to select the inputs for
mutation--using the two metrics induces the same order for most
inputs.
}

\emph{Reproducibility:}
All randomization in TOFU is governed by a random-number seed, optionally under
user control for reproducibility.  However, when running a subject program, TOFU
imposes a wall-clock time limit.  Stopping the execution early can lead to
different basic-block coverage, thereby changing distance measurements,
ultimately changing the results of fuzzing.  We found that multiple TOFU
trials with a given random seed yielded similar results. However, AFLGo and
US-TOFU (which uses AFL's random mutator) form random seeds using system-level
entropy sources, so AFLGo and US-TOFU results can vary more across trials.


\section{Related Work}
\label{Se:Related-Work}

A common criticism of AFL is that it is inefficient at finding bugs deep in a
program~\citep{driller, rawat2017vuzzer,tfuzz, angora}.  Driller~\citep{driller}
uses concolic execution to find deep bugs.  VUZZER~\citep{rawat2017vuzzer} uses
both control-flow and data-flow analysis to decide where and how to mutate
inputs.  T-FUZZ~\citep{tfuzz} removes sanity checks in the source code and then
uses symbolic execution to select valid inputs.
Fairfuzz~\citep{lemieux_fairfuzz:_2018} biases its search in favor of inputs
that execute rarely-executed branches, as those branches are often hard to
cover.  Angora~\citep{angora} uses taint analysis and  gradient descent to solve
the path constraints.  \Citet{Godefroid:2008:GWF:1379022.1375607} use
grammar-based specifications to generate inputs that reach deeper program paths.
TOFU's structured mutator also uses grammars, but differently.  Instead of
generating inputs from scratch, TOFU mutates existing inputs with respect to a
structure specification.  In this way, TOFU can still utilize feedback from
execution on test inputs, as do many AFL-like fuzzers.

As pointed out in~\cite{fuzzingbook2019:index}, the command-line
arguments and options of a program have an important influence on a
program's behavior.
\citep{fuzzingbook2019:index} provides a tool to test command-line arguments and options.
However, it applies only to Python programs, and is a standard fuzzer
(i.e., not target-oriented).
It will not provide the appropriate combination of arguments and
options to reach a particular target location in the program.
In contrast, the generator for command-line languages that TOFU 
provides is general-purpose, and can support most command-line languages
of Unix utilities.
A generated command-line fuzzer is also set up to perform directed
fuzzing, and thus can generate an appropriate combination of
arguments and options to reach---or get close to---a desired target (or
set of targets).

Most fuzzers try to maximize program coverage, but a few have been driven by
other goals.  TIFF~\citep{Jain:2018:TUI:3274694.3274746} specifically focuses on
memory-corruption bugs and adds type inference to the input so that the mutation
to certain bytes is driven by specific type information.
Singularity~\citep{Wei:2018:SPF:3236024.3236039} uses fuzzing to solve the
worst-case-complexity problem by transforming the complexity-testing problem to
an optimal program-synthesis problems and then performing feedback-guided
optimization.  SLOWFUZZ~\citep{slowfuzz} applies evolutionary guidance to
generate inputs that trigger worst-case performance.
RAZZER~\cite{jeong2019razzer} aims to find race-condition bugs in the kernel by
identifying an over-approximation of points at which data races potentially
occur, and guides fuzzing by using a pre-defined system-call grammar and
thread-interleaving tools to trigger data races.  MoonShine~\cite{217573} works
on optimizing the state-of-the-art kernel fuzzer, syzkaller~\cite{syzkaller}.
TOFU also uses fuzzing as a technique to address a non-coverage-maximization problem. TOFU
seeks inputs that reach specific targets in the program, which might
also be tackled using symbolic execution and constraint solving.
TOFU's approach to reach certain targets in a program could make other tools more
efficient.  For example, TOFU could guide RAZZER's subroutine to reach the
locations of potential data races.
Additionally, inputs for triggering worst-case-performance are in general beyond the parsing
stage, so a structured mutator could improve fuzzers' efficiency as
well.

The idea of guided execution is also found in model checking and symbolic
execution~\citep{SAS:DSE, hutchison_high-coverage_2012, su_combining_2015,
Edelkamp:2009:SDM:1530522.1530527, Marinescu:2013:KHT:2491411.2491438}.
\Citet{groce_heuristics_2004} describe different heuristics for model checking
to deal with state-space explosion, including various search and structural
heuristics, which are used to decide what to explore next.  This work is useful
for us because we can likewise apply different heuristics when we generate the input.
However, instead of model checking, which scales poorly with program size, we use guided fuzzing, which may be better suited to handle larger
programs.~\citet{Saxena:2009:LSE:1572272.1572299} devised a novel extension of symbolic execution to analyze input that influences loop executions. Their work addresses the flag and option generation problem. In our work, instead of symbolic execution, we used structured mutation and distance guidance, and do not rely on constraint solvers, which can have scalability problems.  


\section{Conclusion}
\label{Se:Conclusion}

TOFU shows that both structured mutation and distance-based guidance
are beneficial for generating inputs for reaching specific locations
in the program, especially for those deep in the program.
Moreover, TOFU's guidance metric has an intuitive interpretation as a
distance, so users can understand each input's ``closeness'' to the
target.
Our experiments also revealed that ``shallow'' targets, such as ones
in input-parsing code, also commonly appeared during the
software-development process.
Therefore, whether a user chooses either TOFU or AFLGo depends what
kind of target they need to reach.
An AFLGo user could still benefit from TOFU by using TOFU's phase 1 fuzzing
to select flags and options.
AFLGo might also benefit from adopting TOFU's distance metric (and implementation
of the distance computation).


\begin{acks}                            
Supported, in part,
by a gift from Rajiv and Ritu Batra;
by \grantsponsor{GS100000002}{AFRL}{https://afresearchlab.com/} under DARPA MUSE award
  ~\grantnum{GS100000002}{FA8750-14-2-0270} and DARPA STAC award
  ~\grantnum{GS100000002}{FA8750-15-C-0082};
and by \grantsponsor{GS100000003}{ONR}{https://www.onr.navy.mil/} under grants
  ~\grantnum{GS100000003}{N00014-17-1-2889} and 
  ~\grantnum{GS100000003}{N00014-19-1-2318}.
The U.S.\ Government is authorized to reproduce and distribute
reprints for Governmental purposes notwithstanding any copyright
notation thereon.
Opinions, findings, conclusions, or recommendations
expressed in this publication are those of the authors,
and do not necessarily reflect the views of the sponsoring
agencies.
\end{acks}

\bibliography{tofu}


\end{document}